\documentclass[english,twocolumn,showpacs,preprintnumbers,prl]{revtex4}
\usepackage[T1]{fontenc}
\usepackage[latin1]{inputenc}
\usepackage{babel}
\usepackage{graphics}

\makeatletter

\providecommand{\LyX}{L\kern-.1667em\lower.25em\hbox{Y}\kern-.125emX\@}

\makeatother
\begin{document}

\title{{\normalsize The stochastic process of equilibrium fluctuations,
of a system with long range interactions}}

\author{Freddy Bouchet}

\address{Dipartimento di Energetica, Universit{\`a} di Firenze, via S. Marta,
3, 50139 Firenze, Italy}

\date{\today}

\begin{abstract}
The relaxation towards equilibrium of systems with long range interactions
is not yet understood. As a step towards such a comprehension, we
propose the study of the dynamical equilibrium fluctuations in a model
system with long range interaction. We compute analytically, from
the microscopic dynamics, the autocorrelation function of the order
parameter. From this result, we derive analytically a Fokker Planck
equation which describes the stochastic process of the impulsion of
a single particle in an equilibrium bath. The diffusion coefficient
is explicitly computed. 
\end{abstract}

\maketitle

A number of physical systems are governed by long range interactions.
Some examples are given by self-gravitating systems, two dimensional
incompressible, or geophysical flows, some models in plasma physics.
For such Hamiltonian systems, the non additivity of the interactions
makes the usual thermodynamic limit \( N\rightarrow \infty ,V\rightarrow \infty  \)
irrelevant. Microcanonical average is however still relevant, and
generically leads to a mean field description of the equilibrium,
exact in the limit \( N\rightarrow \infty  \) \cite{Bouchet_Barre}.
The relaxation toward equilibrium of these systems still has to be
completely understood. The phenomenology of the dynamics shows that
a rapid relaxation leads to the formation of quasi-stationary structures,
which may be out of equilibrium states (see \cite{Robert_Rosier}
for astrophysical and geophysical examples and \cite{Hors_Equilibre_HMF}
for spin system ones). In most of cases, this is explained by the
existence of stable stationary states of the associated Vlasov equation,
which describes the dynamics by approximating the potential by a mean
field one. In such stable situations, the Vlasov dynamics is a good
approximation of the particle dynamics, on typical time scales diverging
with the number of particle \cite{Braun_Hepp}. The relaxation towards
equilibrium of these structures is then associated to the fluctuations
of the potential around its mean field approximation, and is thus
very slow. One of our goal is to understand such a relaxation, which
is of particular interest, for instance in the study of astrophysical
structures, turbulence parameterization in geophysical flows, etc.
Some works towards a kinetic description of this relaxation have been
proposed, for instance, by Chandrasekar in the context of self-gravitating
systems \cite{Chandrasekar}, Chavanis for the point vortex model
\cite{Chavanis} or for the two-dimensional Euler equation \cite{Chavanis},
or in plasma physics \cite{Esacande_Elskens}. In each of these cases,
the relaxation is then described by a Fokker-Planck equation or some
generalizations. These results are mainly obtained by formal considerations
and the diffusion coefficient in the Fokker-Planck equation is always
expressed as a Kubo formulae or, equivalently, by a formal integral
of the Liouvillian of the dynamics. The diffusion coefficient has
been computed, in some limits, for the point vortex model \cite{Chavanis}
and for self-gravitating systems. 

In the kinetic theory of dilute gases, the Boltzmann equation has
lead to the computation of transport coefficients \cite{Boltzmann}.
This is an example of explicit computation of a diffusion coefficient
for a system, with a large number of particles. A complete mathematical
proof of this result directly from the Hamiltonian dynamics is however
still to be achieved. The computation of the diffusion coefficient
for the standard map \cite{Lichtenberg} is a classical example, for
a system with a small number of degrees of freedom. On the past decades,
the issue of the link between chaotic Hamiltonian dynamics and diffusive
properties has been addressed on a general framework \cite[ ]{Dorfmann}.
We also note works on the relaxation to equilibrium of a massive piston
in interaction with two out of equilibrium perfect gases \cite{Piston},
which a Vlasov like behavior. 

We will show that the diffusion coefficient for systems with long
range interactions can be computed in the large density limit (\( N\rightarrow \infty  \)
with a fixed volume and renormalized interaction). At statistical
equilibrium, one obtains the mean field description typical for long
range interacting systems. Near the equilibrium, particles have an
integrable motion, perturbed by the fluctuations of the mean field
around its equilibrium value. This leads to the relaxation towards
equilibrium. The self consistent nature of the fluctuations (the mean
field oscillates due to small particle deviations, themselves due
to the mean field fluctuations) is however an essential feature of
this process. 

In order to precise these ideas, we consider a simple toy model of
long-range interacting system: the Hamiltonian Mean Field model (HMF).
In this framework, as a first step towards the study of the relaxation
towards equilibrium, we consider the equilibrium dynamical fluctuations.
We first propose an analytic computation of the autocorrelation function
of the mean field order parameter. From this result, we can derive
a Fokker-Planck equation which describes the stochastic process of
a particle in interaction with a bath of \( N-1 \) particles in equilibrium.
The diffusion coefficient is then explicitly computed, from the microscopic
dynamics. We finally conclude by discussing generalization to out
of equilibrium situations, and more realistic models.\\

The Hamiltonian of the attractive HMF model \cite{HMF} is : \begin{equation}
\label{Hamiltonien}
H=\sum ^{N}_{k=1}\frac{p^{2}_{k}}{2}+\frac{1}{2N}\sum ^{N}_{k,l=1}\left( 1-\cos \left( \theta _{k}-\theta _{l}\right) \right) 
\end{equation}
 Because of its simplicity, a large number of authors have considered
this model and its repulsive counterpart (with the opposite sign for
the potential energy). The HMF model is the {}``harmonic oscillator''
for long range interacting systems. We refer to \cite{HMF_Revue}
for a review. Let us define the magnetization \( {\bf M} \) by \( N{\bf M}=\sum ^{N}_{k=1}e^{i\theta _{k}} \)
(\( {\bf M}=M_{x}+iM_{y} \)). Because the kinetic energy per particle
\( e_{c} \) may be exactly expressed as \( 2e_{c}=2E-1+M^{2} \)
(\( E \) is the energy per particle), and because \( M \) is a simple
sum of \( N \) variables, the computation of the static microcanonical
quantities is straightforward. For instance we obtain the volume of
the shell of the phase space, with energy \( E \): \( \Omega \left( E\right) \propto _{N\rightarrow \infty }\int ^{1}_{0}dM\, B\left( M\right) \exp \left( NS\left( E,M\right) \right)  \)
with the entropy \( S \) given by \( S\left( E,M\right) =C\left( M\right) +\log \left( 2E+M^{2}-1\right) /2 \),
where \( C\left( M\right) =\log \left( I_{0}\left( \psi \left( M\right) \right) \right) -M\psi \left( M\right)  \),
\( I_{0} \) is defined by \( 2\pi I_{0}\left( M\right) \equiv \int ^{2\pi }_{0}d\theta \, \exp \left( M\cos \theta \right)  \),
and \( \psi  \) as the inverse function of \( d\log I_{0}/dM \).
The use of the saddle point method, in the previous integral, shows
that an overwhelming number of configurations have a magnetization
close to the equilibrium value \( M_{e}\left( E\right)  \) defined
by \( \partial S\left( E,M_{e}\right) /\partial M=0 \). This equation
shows that, above the critical energy \( E_{c}=3/4 \), whereas below
\( E_{c} \) a second order phase transition occurs. The density in
the \( \mu - \)phase space (all angles and momenta are projected
on a \( \left( \theta ,p\right)  \) space) may be evaluated as \( f_{E}\left( p,\theta \right) \propto \exp \left( -\beta \left( \mathrm{E}\right) \left( p^{2}/2-M_{e}\left( \mathrm{E}\right) \cos \theta \right) \right)  \),
up to a translation of angles. These results are equivalent to the
canonical ones (see \cite{HMF}). 

In the following we consider only energies greater than the critical
one \( E>E_{c} \). In such a case, the equilibrium is homogeneous
: \( M_{e}=0 \). Then \( \beta =1/\left( 2E-1\right)  \) and \( f\left( p,\theta \right) \propto \exp \left( -\beta p^{2}/2\right)  \).
The static fluctuations of \( M \) may also be computed, from the
second derivative of \( S \) with respect to \( M \), at the equilibrium
point. We obtain a Gaussian magnetization with \( N\left\langle {\bf M}^{\star }{\bf M}\right\rangle =2/\left( 2-\beta \right)  \)
and \( \left\langle {\bf M}^{\star }{\bf M}^{\star }\right\rangle =\left\langle {\bf M}{\bf M}\right\rangle =0 \)
(\( {\bf M}^{\star } \) is the complex conjugate of \textbf{\( {\bf M} \))}.
The magnetization has typical fluctuations of order \( N^{-1/2} \),
we thus re-scale it accordingly, by defining :\begin{equation}
\label{Magnetisation}
{\bf m}=\frac{1}{\sqrt{N}}\sum ^{N}_{k=1}e^{i\theta _{k}}.
\end{equation}

The aim of this letter is to study the dynamical equilibrium fluctuations
of this system. From the Hamiltonian (\ref{Hamiltonien}), one obtains
the equations of motion:\begin{equation}
\label{Dynamique}
\frac{d\theta _{k}}{dt}=p_{k}\, \, \, \textrm{and}\, \, \, \frac{dp_{k}}{dt}=\frac{N^{-1/2}}{2}\left( i{\bf m}^{\star }\left( t\right) e^{i\theta _{k}}+{\emph CC}\right) 
\end{equation}
(\( {\emph CC} \) means the complex conjugate of the previous expression).
From the motion equation (\ref{Dynamique}), thanks to the smallness
of the mean field fluctuations, the motion of any particle may be
treated perturbatively in the limit \( N\rightarrow 0 \). We expand
the variables in power of \( N^{-1/2} \) : \( \theta _{k}=\theta _{k,0}+N^{-1/2}\theta _{k,1}+... \),
\( p_{k}=p_{k,0}+N^{-1/2}p_{k,1}+... \) and \( {\bf m}={\bf m}_{0}+N^{-1/2}{\bf m}_{1}+... \)
(the magnetization (\ref{Magnetisation}) has to be self-consistent).
The zero order motion is a free ballistic one : \( p_{k,0}\left( t\right) =p^{0}_{k} \)
and \( \theta _{k,0}\left( t\right) =\theta ^{0}_{k}+p^{0}_{k}t \),
where \( \theta ^{0}_{k} \) and \( p^{0}_{k} \) are the values of
\( \theta  \) and \( p \) for \( t=0 \). The expression (\ref{Dynamique})
clearly shows, that such a perturbative description, around this simple
zero order dynamics, will remain valid as soon as \( t\ll N^{1/2} \).
This expansion leads, to the first order, to \( \theta _{k,1}\left( t\right) =\int ^{t}_{0}du\, p_{k,1}\left( u\right)  \)
with :\begin{equation}
\label{moments_Ordre1}
p_{k,1}\left( t\right) =\int ^{t}_{0}du\, \left[ \frac{i}{2}{\bf m}^{\star }_{0}\left( u\right) e^{i\left( \theta ^{0}_{k}+p^{0}_{k}u\right) }+{\emph CC}\right] .
\end{equation}
A peculiarity of this asymptotic expansion, is that the magnetization
\( {\bf m} \) (\ref{Magnetisation}) is a sum of \( N \) variables,
where \( N^{-1/2} \) is the expansion parameter. A sum of \( N \)
order \( N^{-1/2} \) terms may be of order \( 1 \). To obtain the
zero order magnetization, we thus have to include the first order
expression of the angles \( \theta _{k,1} \). We then obtain:{\footnotesize \begin{equation}
\label{m_0_trajectoire}
{\bf m}_{0}\left( t\right) =a-1/2\int ^{t}_{0}du\int ^{u}_{0}dv\, \left[ b{\bf m}^{\star }_{0}\left( v\right) -c{\bf m}_{0}\left( v\right) \right] 
\end{equation}
}{\small where \( a(t,\theta ^{0}_{k},p^{0}_{k})=N^{-1/2}\sum ^{N}_{k=1}e^{i\left( \theta ^{0}_{k}+p^{0}_{k}t\right) } \),
\( b(t,v,\theta ^{0}_{k},p^{0}_{k})=N^{-1}\sum ^{N}_{k=1}e^{i\left( 2\theta ^{0}_{k}+p^{0}_{k}\left( t+v\right) \right) } \)
and \( c(t,v,\theta ^{0}_{k},p^{0}_{k})=N^{-1}\sum ^{N}_{k=1}e^{ip^{0}_{k}\left( t-v\right) } \).
The} expression (\ref{m_0_trajectoire}) clearly reflects the self-consistent
nature of the motion : the magnetization at time \( t \) depends
on the magnetization at previous times.

From (\ref{m_0_trajectoire}), we compute the magnetization autocorrelation
function at leading order \( \left\langle {\bf m}_{0}\left( t\right) {\bf m}_{0}\left( 0\right) \right\rangle  \),
where the bracket denotes microcanonical averages on the variables
{\footnotesize \( \left( \theta ^{0}_{k},p^{0}_{k}\right)  \)}. We
first note that \( b \) and \( c \) are equal to their microcanonical
average plus fluctuations of order \( N^{-1/2} \). These fluctuations
can be neglected at the order considered. \( b \) and \( c \) are
thus treated as independent of the magnetization. Some lengthy computations
lead to {\footnotesize \( \left\langle a(t,\theta ^{0}_{k},p^{0}_{k}){\bf m}_{0}\right\rangle =2\exp \left( -t^{2}/2\beta \right) /\left( 2-\beta \right)  \)},
{\footnotesize \( \left\langle b(t,v,\theta ^{0}_{k},p^{0}_{k})\right\rangle =0 \)}
and {\footnotesize \( \left\langle c(t,v,\theta ^{0}_{k},p^{0}_{k})\right\rangle =2\exp \left( -\left( t-v\right) ^{2}/2\beta \right) /\left( 2-\beta \right)  \)}
(up to order \( N^{-1/2} \) corrections). From (\ref{m_0_trajectoire}),
we then obtain {\footnotesize \( \left\langle {\bf m}^{\star }\left( t\right) {\bf m}(0)\right\rangle =2\phi \left( t\right) /\left( 2-\beta \right)  \)}
and {\footnotesize \( \left\langle {\bf m}^{\star }\left( t\right) {\bf m}^{\star }(0)\right\rangle =\left\langle {\bf m}\left( t\right) {\bf m}(0)\right\rangle =0 \)};
where the function \( \phi  \) is given by the solution of the integral
equation :\begin{equation}
\label{Integrale_Convolution}
\phi \left( t\right) =\exp \left( -\frac{t^{2}}{2\beta }\right) +\frac{1}{2}\int ^{t}_{0}dv\, v\exp \left( -\frac{v^{2}}{2\beta }\right) \phi \left( t-v\right) 
\end{equation}
 We remark that the rhs integral is a convolution. This makes the
solution of this equation by a Laplace transform natural. We do not
report the result. Whereas the first term on the rhs of this integral
equation is due to the integrable ballistic motion of the particles,
the second term reflects the self-consistent nature of the dynamics.
To our knowledge, it is the first derivation of such a memory term,
directly from the dynamics, in an Hamiltonian system with a large
number of particles. 

\begin{figure}
{\centering \resizebox*{0.3\textwidth}{!}{\includegraphics{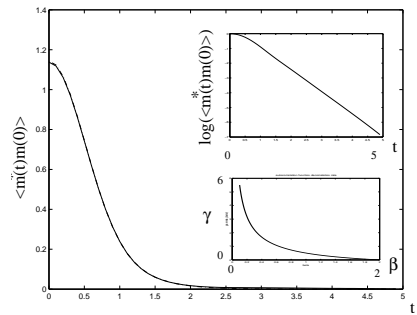}} \par}

\caption{\label{fig:autocorrelation}The magnetization autocorrelation function:
the predicted value (\ref{Integrale_Convolution}) and the numerically
computed value are both represented. They are indistinguishable (maximum
absolute error of \protect\( 3.10^{-3}\protect \)). We have used
\protect\( E=2.5\protect \), \protect\( \beta =1/\left( 2E-1\right) \protect \),
\protect\( N=10\, 000\protect \), and averaged over 18 samples, each
one of duration \protect\( t=8000\protect \). The upper inset shows
that the exponential decay of the autocorrelation function is a good
approximation times greater than \protect\( 2\protect \) or \protect\( 3\protect \),
for these parameters. The lower inset shows the relaxation constant
as a function of the inverse temperature \protect\( \gamma \left( \beta \right) \protect \)
(see \ref{gamma}). }
\end{figure}

To have a physical insight on this autocorrelation function, we compute
the asymptotic behavior of \( \phi  \). Firstly \( \phi \left( t\right) \propto _{t\rightarrow 0}\exp \left( -\left( 2-\beta \right) t^{2}/\left( 4\beta \right) \right)  \).
This approximation is obtained using the value of the second derivative
of \( \phi  \) in \( 0 \) : \( \phi _{1} \). From \ref{Integrale_Convolution},
by a Taylor expansion, we compute \( \phi _{1}=\left( 2-\beta \right) /\left( 2\beta \right)  \).
Such a Gaussian behavior for small times, would be typical of a ballistic
behavior. However, we note that the coefficient \( \left( 2-\beta \right) /\left( 2\beta \right)  \)
is not uniquely due to the integrable zero order dynamics, but is
renormalized by the memory term. Secondly we obtain : {\small \begin{equation}
\label{gamma}
\phi \left( t\right) \propto _{t\rightarrow \infty }\mathrm{A}\left( \beta \right) \exp \left( -\gamma \left( \beta \right) t\right) \, \, \rm {;}\, \, \gamma \left( \beta \right) =\left( 2/\beta \right) ^{1/2}F^{-1}\left( \beta \right) ,
\end{equation}
}where \( F^{-1} \) is the inverse of the function \( F \), with
\( F\left( x\right) =2/\left( 1+\sqrt{\pi }x\exp \left( x^{2}\right) \rm {erfc}\left( -x\right) \right)  \),
where \( \rm {erfc} \) is the complementary error function This exponential
limit for the autocorrelation function is natural, as it corresponds
to the Markovian limit for the magnetization stochastic process. Using
the expression (\ref{gamma}) as an Ansatz and evaluating the integral
equation (\ref{Integrale_Convolution}) at dominant order for \( t\rightarrow \infty  \),
it is possible to obtain this result for \( \gamma  \). The lower
inset of Fig. \ref{fig:autocorrelation} shows the relaxation constant
\( \gamma  \) as a function of \( \beta  \). Near the critical energy
(\( \beta _{c}=2 \)), the relaxation constant tends to \( 0 \).
This indicates that near the critical point, the relaxation time diverges.
On the contrary, for large energy, \( \gamma  \) diverges and the
relaxation time is very small. Let us compare these results for the
autocorrelation function with numerical results. We first numerically
compute the theoretical prediction for \( \phi  \) (\( \phi \left( t\right)  \)
may be numerically computed either from its Fourier transform, its
Laplace transform or by an iterative scheme directly from (\ref{Integrale_Convolution})).
Fig. \ref{fig:autocorrelation} shows a comparison of this theoretical
autocorrelation function with the one obtained directly from the integration
of the Hamiltonian dynamics (\ref{Hamiltonien}).

Because the stochastic process is stationary, using the Wiener-Kinchin
theorem, the spectral density of the complex magnetization may be
computed from the Fourier transform of the autocorrelation function.
As the integral equation describing the autocorrelation function is
a convolution, the computation of this spectral density is easy. Defining
\( S\left( \omega \right) =2/\pi \int ^{\infty }_{0}dt\, \cos \left( \omega t\right) \left\langle {\bf m}^{\star }\left( t\right) m{\bf m}0)\right\rangle  \),
one obtains :{\small \begin{equation}
\label{densite_spectrale}
S\left( \omega \right) =\frac{8\left( \pi \beta /2\right) ^{1/2}\exp \left( -\beta \omega ^{2}/2\right) }{2\left[ \left( 2-\beta \right) +\beta ^{3/2}\omega A\left( \beta ^{1/2}\omega \right) \right] ^{2}+\pi \beta ^{3}\omega ^{2}\exp \left( -\beta \omega ^{2}\right) }
\end{equation}
}where {\footnotesize \( A\left( x\right) =\exp \left( -x^{2}/2\right) \int ^{x}_{0}du\, \exp \left( u^{2}/2\right)  \)}.
We have \( S\left( \omega \right) \sim _{\omega \rightarrow \infty }\left( 2\beta /\pi \right) ^{1/2}\exp \left( -\beta \omega ^{2}/2\right)  \). 

\begin{figure}
{\centering \resizebox*{0.2\textwidth}{!}{\includegraphics{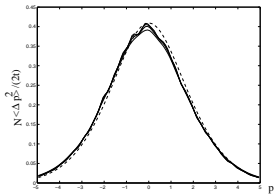}} \par}

\caption{\label{fig:Deltap2}The solid curve shows the mean square displacement
of a particle in function of its initial momentum ,normalized by \protect\( N\protect \)
and divided by the time (\protect\( N\left\langle \Delta p^{2}\right\rangle \left( p,t\right) /\left( 2t\right) \protect \))
(this is not a distribution), for four values of time : \protect\( t=10,15,20,25\protect \)
; \protect\( N=10\, 000\protect \), \protect\( \beta =1/4\protect \)
. As the curves are superposed for time, this shows that the motion
is actually diffusive. The dashed curve represents the predicted result
\protect\( D\left( p\right) \protect \) (equation \ref{densite_spectrale}
with \protect\( D\left( p\right) =\pi S\left( p\right) /4\protect \)),
no fit. This confirms the theoretical analysis, up to errors due to
an incomplete statistics. }
\end{figure}

Let us now consider the diffusion of the momentum \( p \) of a single
particle, where all others particles have a random angle and momentum
according to the microcanonical distribution (one particle in a bath
at equilibrium). Let us denote \( \left\langle \Delta p\right\rangle \left( p,t\right)  \)
the mean displacement \( p(t)-p(0) \), and \( \left\langle \Delta p^{2}\right\rangle \left( p,t\right)  \)
the mean square displacement of a particle knowing that its initial
momentum (\( p\left( 0\right) =p \)). From (\ref{moments_Ordre1})
and the results for the autocorrelation function or for similar quantities,
it is possible to compute explicitly \( \left\langle \Delta p^{2}\right\rangle \left( p,t\right)  \),
at the leading order in \( N \), for any time such that \( t\ll N^{1/2} \)
(perturbative description of the dynamics). The quantity \( \left\langle \Delta p^{2}\right\rangle \left( p,t\right)  \)
has a transient behavior on a time scale of order \( 1 \) (the explicit
computation is feasible, but not reported), followed for \( 1\ll t\ll N^{1/2} \),
by a diffusive behavior. We then obtain \( \left\langle \Delta p^{2}\right\rangle \left( p,t\right) \sim _{1\ll t\ll N^{-1/2}}2D\left( p\right) N^{-1}t \),
with \begin{equation}
\label{Diffusion}
D\left( p\right) =\frac{1}{2}\int ^{\infty }_{0}dt\, \left\langle {\bf m}^{\star }\left( t\right) {\bf m}(0)\right\rangle \cos \left( pt\right) 
\end{equation}
This result is the equivalent of a Kubo formulae. However it states
a bite more: the diffusion coefficient is there expressed as the autocorrelation
of the mean field and not as the autocorrelation of the force. We
note that the diffusion coefficient is proportional to the spectral
density : \( D\left( p\right) =\pi S\left( p\right) /4 \). This is
a peculiarity of this model for which the interaction is built with
a cosine. An analytical expression for \( D \) is thus obtained from
(\ref{densite_spectrale}). The computation of \( \left\langle \Delta p\right\rangle \left( p,t\right)  \)
may be done following the same procedure. Please note, however, that
the \( N^{-1/2} \) contribution vanishes. The lower order contribution
comes from a perturbative description of the dynamics at order 2 (\( p_{k,2}\left( t\right)  \)
and \( {\bf m}_{1}\left( t\right)  \)). The systematic momentum change
is \( N\left\langle \Delta p\right\rangle \left( p,t\right) \sim _{1\ll t\ll N^{-1/2}}\left( dD\left( p\right) /dp-\beta pD\left( p\right) \right) t \).
In the following, we will see that this result can be deduced from
the stationarity of the stochastic process (or equivalently from the
fact that the distribution for \( p \) tends to the equilibrium distribution).
This is the equivalent of an Einstein relation. Fig. \ref{fig:Deltap2}
shows that the analytical diffusion coefficient with the numerically
computed \( N\left\langle \Delta p^{2}\right\rangle \left( p,t\right) /t \)
agree.

We have observed a diffusive behavior for the momenta (\ref{Diffusion})
with a systematic momentum drift, for \( 1\ll t\ll N^{1/2} \). Moreover
the mean displacement and the mean square displacement are small as
they scale like \( N^{-1} \). These two facts are the two hypothesis
for the derivation of a Fokker-Planck equation. Thus any momentum
distribution function \( f\left( p\right)  \) evolves, at the leading
order in \( N \), through the equation:\begin{equation}
\label{Fokker-Planck}
\frac{\partial f}{\partial t}=\frac{1}{N}\frac{\partial }{\partial p}\left( D\left( p\right) \left( \frac{\partial f}{\partial p}+\beta pf\right) \right) 
\end{equation}
 This equation is valid for time \( t\gg 1 \). For the derivation
of the mean square and mean displacement, we have assumed \( t\ll N^{1/2} \)
(perturbative description). However, the previous analysis has also
shown that the correlation function decays exponentially for large
time. The correlation time for the force (or equivalently the magnetization)
is then of order \( 1 \) and is thus much smaller than \( N^{1/2} \).
This is a first indication that the stochastic process may becomes
Markovian for time much smaller than \( N^{1/2} \). If it is actually
such, the Fokker-Planck will be correct for any time \( t \). We
note that this equation actually converges towards the equilibrium
density \( P_{eq}\left( p\right) =\left( \beta /2\pi \right) ^{1/2}\exp \left( -\beta p^{2}/2\right)  \). 

In this letter, we have analytically computed the autocorrelation
function for the HMF model. We have used this result to compute analytically
the diffusion of the momentum of a single particle in an equilibrium
distribution. We have obtained a Fokker-Planck equation for which
the diffusion coefficient is explicitly computed. A more complete
study of the magnetization stochastic process, the detailed computations,
and the study of this Fokker-Planck equation will be addressed in
a forthcoming paper \cite{Analyse_Fokker_Planck}. Due to the asymptotic
decay of the diffusion coefficient, for large momentum, the spectrum
of the linear operators of the Fokker-Planck equation has no gap between
the eigenvalue corresponding to the ground state and the other eigenvalues.
In the introduction, we have introduced this work as a first step
towards the description of relaxation towards the statistical equilibrium.
We hope to generalize in the future these results to such out-of equilibrium
states. Let us note however, that the Fokker-Planck equation should
correctly describe the relaxation towards equilibrium of states sufficiently
close to equilibrium. The generalization of this letter results to
other long range interacting particle models may follow the same path,
with technical difficulties linked to the continuous nature of the
mean field in most systems, and the theoretical problems linked with
the divergence of some interactions at small scales (point-vortices,
self-gravitating systems and plasma). 

I acknowledge useful discussion with J. Barré, E. Caglioti, P.H. Chavanis,
T. Dauxois, Y.Elskens, S. Ruffo, A. Vulpiani and Y. Yamaguchi. This
work has been supported by the contract COFIN00 \emph{Chaos and Localization
in Classical and Quantum mechanics} and by the European network \emph{Stirring
and mixing,} RTN2-2001-00285.


\begin{thebibliography}{10}
\bibitem{Bouchet_Barre}F. Bouchet and J.Barr\'e, submitted to J. Stat. Phys, condmat 0303307.
\bibitem{Robert_Rosier}D. Lynden-Bell, Mon. Not. R. Astr. Soc. \textbf{136} 101 (1967) R.
Robert and C. Rosier, J.Stat. Phys \textbf{86} 481 (1997). 
\bibitem{Hors_Equilibre_HMF}V. Latora, A. Rapisarda and C. Tsallis, Phys.Rev.E 64 (2001) 056134.
\bibitem{Braun_Hepp}W. Braun and K. Hepp Comm. Math. Phys. \textbf{56} 101-113 (1977)
H. Spohn, \emph{Large Scale Dynamics of Interacting Particles}, Springer-Verlag,
Heidelberg und Berlin, (1991). 
\bibitem{Chandrasekar}S. Chandrasekar, \emph{Principles of stellar dynamics} (Dover 1942),
S. Chandrasekar, Rev. Mod. Phys. \textbf{21}, 383 (1949). \textbf{}
\bibitem{Chavanis}P.H. Chavanis, in \emph{Dynamics and Thermodynamics of Systems with
Long Range Interactions,} T. Dauxois, S. Ruffo, E. Arimondo, M. Wilkens
Eds., (Lecture Notes in Physics Vol. 602, Springer, 2002). cond-mat/0212223
; P.H. Chavanis, Phys. Rev. Lett. \textbf{84} 5512 (2000) ; P.H. Chavanis
and C. Sire, Phys. Fluids \textbf{13} 7 1804 (2001) ; P.H. Chavanis,
Phys. Rev. E \textbf{63} 065301(R).
\bibitem{Esacande_Elskens}Y. Elskens and D.F. Escande, \emph{Microscopic dynamics of plasmas
and chaos} (IoP Publishing, Bristol, 2002).
\bibitem{Boltzmann}E.M. Lifshitz and L.P. Pitaevskii \emph{Physical kinetics} (1981)
\bibitem{Lichtenberg}A.J. Lichtenberg and M.A. Lieberman, \emph{Regular and chaotic dynamics}
(second edition, Springer Verlag, 1992).
\bibitem{Dorfmann}P. Gaspard, Chaos, scattering and statistical mechanics (1999); B.
Dorfmann, \emph{An introduction to chaos in non-equilibrium statistical
mechanics} (1999)
\bibitem{Piston}N. Chernov and J. Lebowitz, J. Stat. Phys. \textbf{109,} 507 \textbf{}(2002)
; N. Chernov, J. Lebowitz and Ya. Sinai, J. Stat. Phys. \textbf{109,}
529 \textbf{}(2002) ; E. Caglioti, N. Chernov and J. Lebowitz, preprint.
\bibitem{HMF}M. Antoni and S. Ruffo, Phys. Rev. \emph{}E \textbf{52} (3), 2361-2374
(1995). 
\bibitem{HMF_Revue}T. Dauxois, V. Latora, A. Rapisarda, S. Ruffo and A. Torcini, in \emph{Dynamics
and Thermodynamics of Systems with Long Range Interactions,} T. Dauxois,
S. Ruffo, E. Arimondo, M. Wilkens Eds., (Lecture Notes in Physics
Vol. 602, Springer, 2002).
\bibitem{Analyse_Fokker_Planck}F. Bouchet, J. Barré, T. Dauxois and S. Ruffo, to be published (2003). \end{thebibliography}
\end{document}